# Study the vibration and dynamic response of the dipole girder system for CSNS/RCS

Liu Ren-Hong (刘仁洪)[1,2;1]   Wang Min(王敏)[1]   Zhang Jun-Song(张俊嵩)[2]   Wang Guang-Yuan (王广源)[2]

[1] No. 58 Research Institute of China Ordnance Industries, Mianyang 621000, China

[2] Institute of High Energy Physics, Chinese Academy of Sciences, Beijing 100049, China

**Abstract:** China Spallation Neutron Source (CSNS) is a high intensity proton accelerator based facility, and its accelerator complex includes two main parts: an H- linac and a rapid cycling synchrotron (RCS). The RCS accumulates the 80MeV proton beam, and accelerates it to 1.6GeV, with a repetition rate of 25Hz. The dipole of the CSNS/RCS is operated at a 25 Hz sinusoidal alternating current which causes severe vibrate. The vibration will influence the long-term safety and reliable operation of the magnet. The dipole of the CSNS/RCS is an active vibration equipment which is different from the ground vibration accelerator. It's very important to design and study the dynamic characteristics of the dipole girder system. This paper takes the dipole and girder as a specific model system, a method for studying the dynamic characteristics of the system is put forward by combining theoretical calculation with experimental testing. The modal parameters with and without vibration isolator of the dipole girder system are obtain through ANSYS simulation and testing. Then the dynamic response of the system is calculated with modal analysis and vibration testing data. The dipole girder takes four-point support system which maybe appears over-constrained (three-point support) in the progress of adjustment. The dynamic characteristics and dynamic response of the three point girder system were studied with the same method.

## 1 Introduction

The CSNS-I accelerators consist of an 80 MeV H− linac and a rapid cycling synchrotron of 1.6 GeV [1-2]. The RCS ring is a four-folded symmetrical topological structure which consists of four arc zones and four line segments. There are 24 sets of dipole magnets uniformly distributed in the whole RCS ring, and the magnets will be operated at a 25 Hz rate sinusoidal alternating current. The magnetic core and coils will make severe vibration especially at the frequency 25 Hz such as the J-PARC AC dipole. The vibration influenced other equipment through the magnet girder system.

The dipole girder system with complex structure and high precision adjustment is one of the most important equipment of the CSNS/RCS. Because of the self-excited vibration, the comprehensive technical index of requirement is different from other accelerators whose vibration is caused by the ground vibration. So it is necessary to study the dynamic characteristics and reduce the vibration of the system [3-4]. The theoretical modal analysis and testing modal analysis are the main research methods. The theoretical modal analysis is based on the liner vibration theory and finite element method to study the relationship among the excitation, system and response. The testing modal analysis uses the input and response parameters to obtain the modal parameters (frequency, damping ratio and vibration mode) [5]. The dynamic characteristics of the girder are very important. This paper adopts the dipole girder system as the research object. The theoretical and testing methods are used to study the dynamic characteristics of the system. After that a new isolator was designed to improve the dynamic characteristics of the system, decrease the vibratory force and avoid the resonance phenomenon. The girder uses four-point support which can increase the stiffness of dipole-girder system. But the girder maybe appears over-constrained (three-point support) in the progress of adjustment. The dynamic characteristics and dynamic response of the three-point girder system were also important to be studied.

---

[1] Email: liu397574933@126.com
liurh@ihep.ac.cn

## 2 Modal analysis of the dipole-girder system

The theoretical modal analysis and testing modal analysis are the main methods to research the dynamic characteristics of the system. The theoretical modal analysis is based on the liner vibration theory and finite element method to research the relationship among the excitation, system and response. This paper uses ANSYS software to simulate the dipole girder system with and without vibration isolator. The finite element structure (FE) is consists of dipole, girder. The dipole is composed of silicon steel sheets, steel plate, stainless plate and aluminum coil, and the girder is composed of steel plate and adjusting mechanism. The FE is constructed with element of solid-186, and the damping material is constructed with element of combin-40. In the progress of analysis, the Block Lanczos Method is used to calculate the natural frequency and vibration mode of the system. The top six natural frequencies are obtained as shown in Table 1.

Table 1. The modal analysis simulation results of the dipole-girder system

| Modal order | 1 | 2 | 3 | 4 | 5 | 6 |
|---|---|---|---|---|---|---|
| ANSYS natural frequency without isolator f/Hz | 6.658 | 11.23 | 14.59 | 23.77 | 28.45 | 36.33 |
| ANSYS natural frequency with isolator f/Hz | 5.984 | 10.289 | 14.113 | 18.789 | 22.741 | 26.165 |

At the same time the testing modal analysis obtain the natural frequency parameters of the system with the curve fitting analyses of the transfer function of the structure's excitation and response (such as acceleration, velocity, displacement, etc.). [7-8]. In testing modal, the transfer function can be calculated from the exciting point and responding point parameters. The different order modal parameters can be calculated from any one row or one column of elements. This testing modal analysis takes force hammer excitation system. The modal parameters identification method of MIMO is taken too. There are 60 measuring points arranged around the whole system according to the selecting principle, 24 points arranged on the dipole to measure the X, Y and Z direction acceleration of the 12 corner points, and 36 points arranged on the girder to measure the three direction acceleration of the first and third plate's corner points. The accelerator sensor arrangement and testing method of with vibration isolator are the same as the system without vibration isolator. The testing system and the acceleration sensor arrangement are shown in Fig. 1.

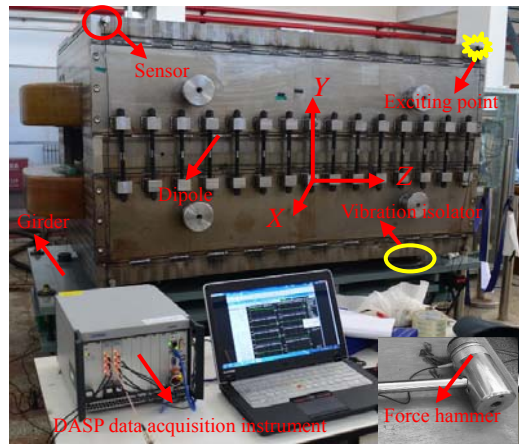

Fig.1. The experiment layout of the testing modal

The force hammer and vibration response signals are acquired by the intelligent analyzer of INV3032C. After testing, the system testing modal parameters (natural frequency and damping ratio) have been got through the data processing analysis system of DASP. The testing results are show in Table 2.

Table 2. The testing modal analysis results of the dipole-girder system

| Modal order | 1 | 2 | 3 | 4 | 5 | 6 |
|---|---|---|---|---|---|---|
| Testing natural frequency without isolator f/Hz | 8.66 | 11.09 | 15.33 | 23.21 | 28.44 | 36.22 |
| Testing damping ratio without isolator /% | 1.772 | 2.602 | 2.647 | 5.519 | 2.786 | 3.922 |
| Testing natural frequency with isolator f/Hz | 7.25 | 10.47 | 15.48 | 16.81 | 22.34 | 26.38 |
| Testing damping ratio with isolator /% | 1.727 | 2.101 | 2.196 | 9.293 | 3.264 | 2.845 |

Fig.2 shows the natural frequencies contrast between simulations and testing. The two conditions simulation and testing almost have the same modal shapes, and both the fourth modal shape is a vertical movement at X-Z plane (Y direction). With the vibration isolator the high order natural frequencies are decreased. Contrasting the natural frequencies of the two conditions which shows the fourth natural frequency with vibration isolator is decreased to 16.81 Hz. And the damping ratio was increase to 9.293% which is close to the design objective 10%. The modal analysis results indicate the dipole-girder would not take resonance phenomenon. The simulation results are almost identical with the testing results. So the modal analysis of the system and the FE is reasonable.

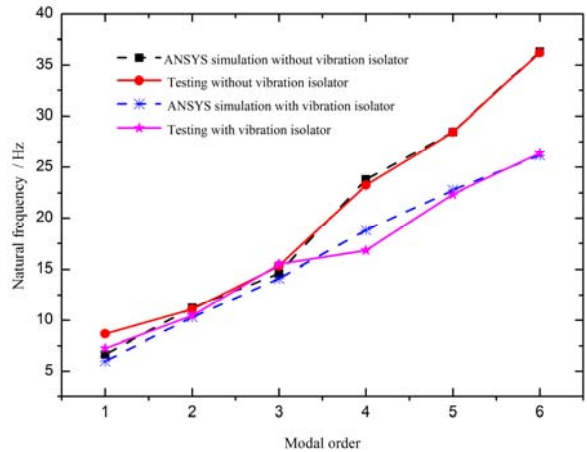

Fig.2. The natural frequencies contrast of the girder

## 3 Dynamic response analysis of the dipole-girder system

The traditional structural strength analysis is static analysis. The static analysis results usually satisfy the material property. The maximum static strain and maximum static stress of the system is 0.042mm and 33.4MPa. The maximum stress is less than the permissible stress 345MPa of material. It is a safe system if the dipole is not vibrating. Because of the safe-excited vibration, it is necessary to research the dynamic characteristics response of the system. The dynamic response of the system can be simulated through ANSYS. The vibration testing data will be the excitation signal of the simulation. The dynamic characteristics response is calculated with modal analysis and vibration testing data, and the computational procedure which shows in Fig.3.

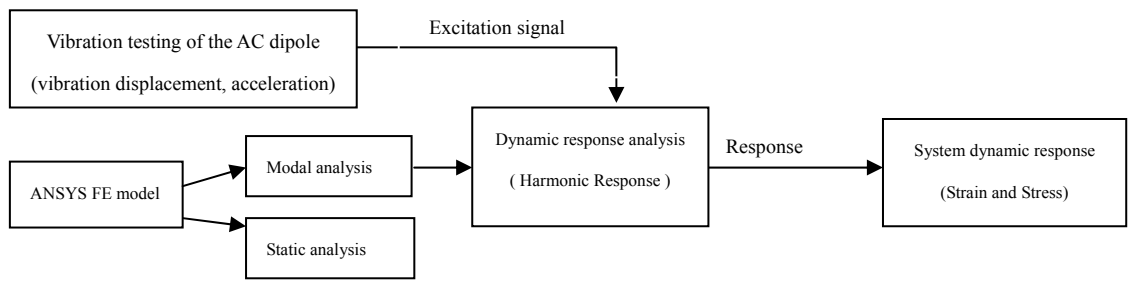

Fig.3. The computational procedure of the dynamic analysis

In this paper a new vibration absorber is designed for the CSNS/RCS dipole with special material. Then the acceleration sensor will be used to measure the vibration of the dipole. The average acceleration magnitude of the dipole will be the excitation source of the vibratory force. And the main frequency of the force is set as 25 Hz. With the harmonic response simulation method, the dynamic response of the dipole-girder system were calculated. The dynamic strain and dynamic stress were obtained as shown in Fig. 4, Fig. 5 and Table 3. And the harmonic response curves of the dipole were shown in Fig. 6 and Fig. 7.

Table 3. The dynamic strain and stress of the dipole-girder system

| Simulation situation | Without vibration isolator | With vibration isolator |
|---|---|---|
| Dynamic strain(mm) | 0.064 | 0.026 |
| Dynamic tress(MPa) | 42.65 | 20.64 |

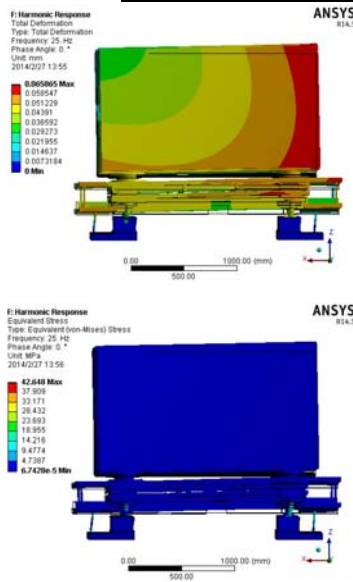

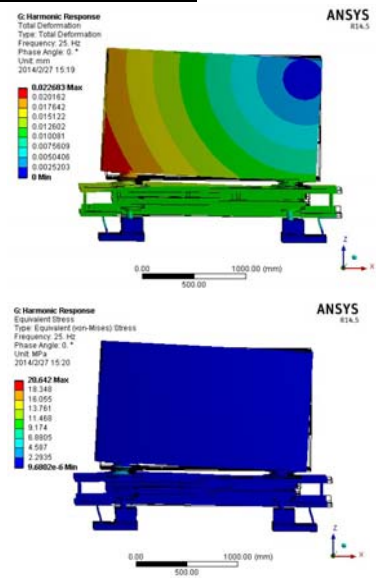

Fig.4. The dynamic strain and stress analysis without vibration isolator

Fig.5. The dynamic strain and stress analysis with vibration isolator

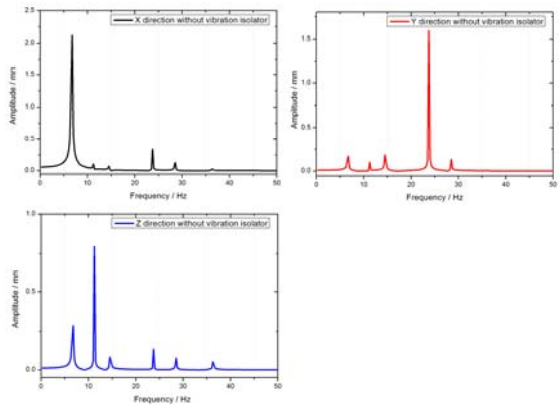

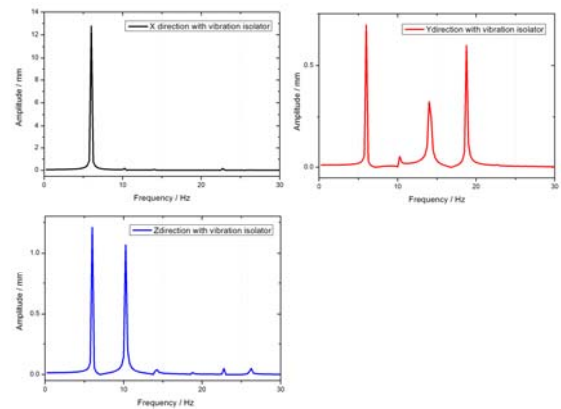

Fig.6. The harmonic response curves of the dipole without isolator

Fig.7. The harmonic response curves of the dipole with isolator

The simulation results explain the dynamic response analysis is quite different with the traditional static analysis. The maximum dynamic strain is more than the static strain without vibration isolator. The dynamic strain maybe cause strain fatigue in the long-term operation. With the vibration isolator, the dynamic strain and stress is less than the system without vibration isolator. And the vibration isolator also changes the dynamic stress's position. The maximum dynamic stress is located at the vertical adjustment system without vibration isolator. The maximum dynamic stress is located at the spring with vibration isolator. The vibration isolator improves the dynamic characteristics of the system. With the vibration isolator the dynamic strain and stress reduce half which will decrease the strain and stress fatigue odds of the weld reinforcing steel plate. At the same time the vibration harmonic response characters of the dipole are smaller with isolator, and there is no resonance peak around the exciting frequency 25Hz. The system response spectrum of the dipole is also not in resonance interval, so the vibration isolation system achieved the expected effect.

# 4 The unstable analysis of the dipole girder system

The dipole of the CSNS/RCS has strong self-excited vibrate, so the stability performance of the girder is particularly important. The dipole girder takes four-point support scheme which can improve the stability and stiffness of dipole girder system of the CSNS/RCS. The brief four-point support scheme of the girder is showed in Fig. 8. As is known to all three-points can form a plane, so the dipole girder is an over-constrained girder which may decrease the stability of the dipole girder system. The girder maybe arise three-point support form in the progress of adjusting. The three-point support forms are BCD ABC and ACD. This paper takes the dipole girder with vibration isolator as a specific model system, the unstable analysis of the dipole girder system are studied through the structure dynamic characteristics. The top six natural frequencies are obtained as shown in Table 4. The three-point support forms have the same modal vibration mode which shows the whole rock of the dipole girder system.

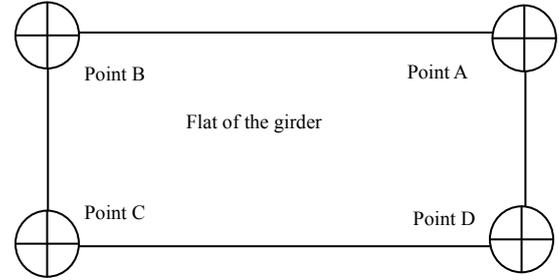

Fig.8. The brief four-point support scheme of the girder

Table 4. The modal analysis simulation results of the dipole-girder system with three point support forms

| Modal order / Support form | 1 | 2 | 3 | 4 | 5 | 6 |
|---|---|---|---|---|---|---|
| BCD | 3.7346Hz | 6.1967 Hz | 10.55 Hz | 14.295 Hz | 18.948 Hz | 22.676 Hz |
| ABC | 3.7321 Hz | 6.1698 Hz | 10.516 Hz | 14.189 Hz | 18.727 Hz | 22.591 Hz |
| ACD | 3.4818 Hz | 7.5401 Hz | 9.9443 Hz | 14.902 Hz | 19.587 Hz | 21.965 Hz |

Fig. 9 shows the natural frequencies contrast between normal support (four-point support) and three-point support. The natural frequencies of the three-point support system are smaller than the normal system. Especially the top three order close to 1 times the amount of reduced. So the three-point support form reduces the stiffness vibration resistance and stability of the dipole girder system. Compared with the normal support, the vibration modal shapes of the three-point support have changed which the third and fourth order modal shape exchanged.

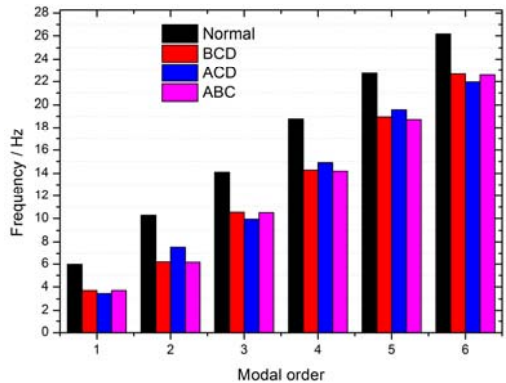

Fig.9. The natural frequencies contrast of the normal and three-point support girder

This paper studies the harmonic response characters of three-point support dipole girder system. And the contrast figure is shown in Fig. 10. The harmonic response curves of the three kinds of three-point support form are basically identical. The three-point support system are obvious magnified the response amplitude of the dipole, especially at the resonance frequency response amplitude. The high frequency vibration frequency response of the three-point support dipole girder system is very small. But the low frequency of 3.7Hz dynamic response amplitude is very large which will reduces the horizontal stability of the system. So the three-point support form will seriously affect the running stability of the dipole girder system which maybe cause stress uneven and produce larger local dynamic strain and stress.

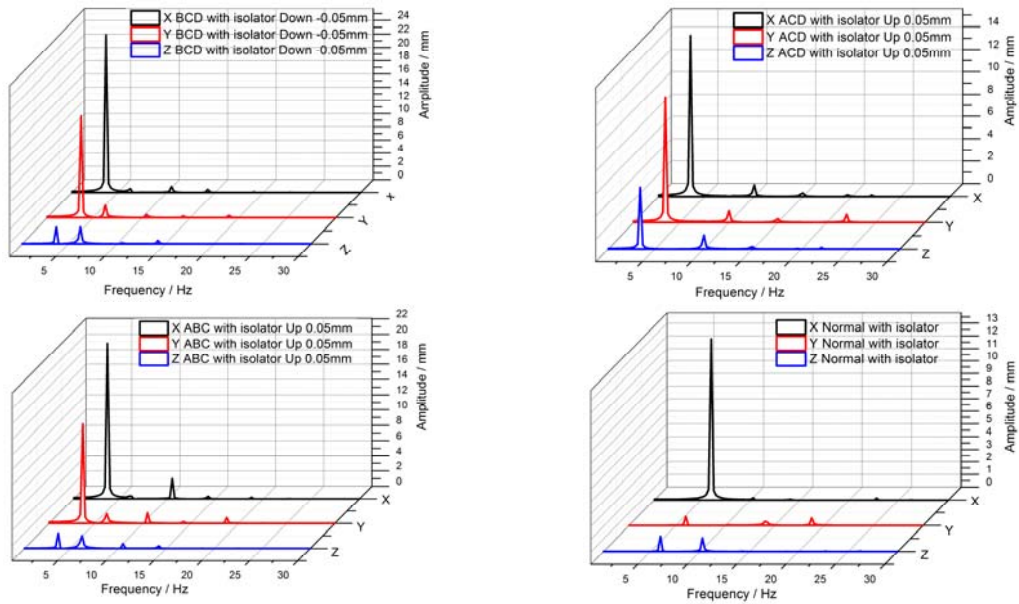

Fig.10. The harmonic response curves of the normal and three-point support girder

## 5 Conclusions

The dipole girder system play a very important role in the accelerator of CSNS/RCS, so studying the vibration of the system is necessary. This paper established the suitable finite element structure of the dipole girder system, a method for analysing and studying the dynamic characteristic of the system is put forward by combining theoretical calculation (ANSYS simulation) with experimental testing. So the resonance phenomenon can be avoided and the structure vibration resistance can be improved before manufacture. A new isolator was designed to decrease the vibration amplitude and the vibratory force transmission of the dipole. In order to improve the vibration resistance and stability of system, the dipole girder system of CSNS/RCS take four-point support form. The dipole girder system is likely to appear three-point support form in the actual process of adjustment which will reduce the stability and vibration resistance of the system. So it is important to optimize the adjustment scheme of dipole girder to avoid three-point support form appears which can ensure the long-term operation stability of the dipole girder system. The active vibration of magnet is different with passive vibration which was caused by ground vibration. So this paper can provide a reasonable way to design and research the equipment which has self-excited vibrate such as AC dipole, AC quadrupole [9] and superconductor cavity system of CSNS, etc.

*The authors would like to thank other CSNS colleagues to give help and advice.*